\documentclass[aps,prl,superscriptaddress,twocolumn,showpacs,dvipdfmx]{revtex4-1}
\usepackage{graphicx}
\usepackage{latexsym}
\usepackage{fancybox}
\usepackage{color}

\def\vector#1{\mbox{\boldmath $#1$}}

\begin{document}
\title{$T/B$ scaling without quasiparticle mass divergence: YbCo$_2$Ge$_4$}
\author{Akito Sakai}
\affiliation{Experimental Physics VI, Center for Electronic Correlations and Magnetism, University of Augsburg, Germany}
\author{Kentaro Kitagawa}
\affiliation{Department of Physics, University of Tokyo, Tokyo 113-0033, Japan}
\author{Kazuyuki Matsubayashi}
\affiliation{Department of Engineering Science, University of Electro-Communications, Chofu, Tokyo 182-8585, Japan }
\author{Makoto Iwatani}
\affiliation{Graduate School of Integrated Arts and Sciences, Kochi University, Kochi 780-8520, Japan}
\author{Philipp Gegenwart}
\affiliation{Experimental Physics VI, Center for Electronic Correlations and Magnetism, University of Augsburg, Germany}
\date{\today}

\begin{abstract}
YbCo$_2$Ge$_4$ is a clean paramagnetic Kondo lattice which displays non-Fermi liquid behavior. We report a detailed investigation of the specific heat, magnetic Gr\"uneisen parameter ($\Gamma_{\rm mag}$) and temperature derivative of the magnetization ($M$) on a high-quality single crystal at temperatures down to $0.1$~K and magnetic fields up to 7~T. $\Gamma_{\rm mag}$ and $dM/dT$ display a divergence upon cooling and obey $T/B$ scaling. Similar behavior has previously been found in several other Yb-based Kondo lattices and related to a zero-field quantum critical point without fine tuning of pressure or composition. However, in the approach of $B\rightarrow 0$ the electronic heat capacity coefficient of YbCo$_2$Ge$_4$ saturates at low $T$, excluding ferromagnetic quantum criticality. This indicates that $T/B$ scaling is insufficient to prove a zero-field quantum critical point.
\end{abstract}

\maketitle
%\section{Introduction}
Zero-temperature continuous phase transitions, called quantum critical points (QCPs) have been discussed in the context of non-Fermi liquid (NFL) behavior or unconventional superconductivity \cite{Mathur1998}. Ce- or Yb-based Kondo lattices, whose ground state sensitively depends on the balance between on-site Kondo screening and intersite magnetic couplings, are prototype materials to study quantum criticality. QCPs have been realized experimentally in several 4$f$-electron based compounds by composition, pressure or magnetic field tuning \cite{Loehneysen2007}. However, there is an increasing number of materials that are suggested to display quantum criticality upon cooling without tuning any of these parameters. Prominent examples include CeNi$_2$Ge$_2$~\cite{RKuchler2003}, CeRhSn \cite{Tokiwa2015}, CeCoIn$_5$ \cite{Tokiwa2013}, Pr$_2$Ir$_2$O$_7$~\cite{Tokiwa2014}, $\beta$-YbAlB$_4$~\cite{Nakatsuji, Matsumoto,Tomita2015} and the quasicrystal Au$_{51}$Al$_{34}$Yb$_{15}$~\cite{Deguchi}.

It is very unlikely, that a compound is accidentally located at such a special point in multidimensional phase space. Chemical substitution studies on CeNi$_2$Ge$_2$ indeed suggest that this compound is actually located slightly beyond the QCP on the paramagnetic side \cite{Fukuhara2002}, though the crossover between quantum critical and Fermi liquid behavior in the undoped material is too low in $T$ to be detectable. Relatedly, for CeCoIn$_5$ the QCP is masked by superconductivity and the conclusion of zero-field quantum criticality relies on extrapolation~\cite{Tokiwa2013}. On the other hand, for CeRhSn~\cite{Tokiwa2015} and Pr$_2$Ir$_2$O$_7$~\cite{Tokiwa2014} zero-field quantum criticality originates from strong geometrical frustration which effectively suppresses long-range order. An even more exotic scenario would be the existence of an extended quantum critical regime. Such a quantum critical phase may be sensitive to the application of magnetic fields but stable over a substantial range of applied pressure. Evidence for pressure insensitivity of NFL behavior has been reported for $\beta$-YbAlB$_4$~\cite{Tomita2015} and Au$_{51}$Al$_{34}$Yb$_{15}$~\cite{Deguchi}. However, the nature of a quantum critical phase, instead of singlar QCP, remains unclear. 

In this paper, we discuss a new Yb-based Kondo lattice, YbCo$_2$Ge$_4$~\cite{Kitagawa}, which displays similar temperature over magnetic field, $T/B$ scaling behavior as found previously for $\beta$-YbAlB$_4$~\cite{Nakatsuji, Matsumoto} and the quasicrystal Au$_{51}$Al$_{34}$Yb$_{15}$. Note, that in contrast to the case of field-induced QCPs~\cite{LZhu2003,MGarst2005}, it involves the applied field $B$ and not a tuning parameter $r=(B-B_c)/B_c$, because the critical field $B_c=0$. We observe a divergence of the magnetic Gr\"uneisen parameter and the temperature derivative of the magnetization $dM/dT$ upon cooling in the limit of zero field similar as in the above mentioned examples. However,
the heat capacity coefficient is almost constant in the respective $T$-$B$ range, excluding quantum criticality as origin of $T/B$ scaling.

Our material of interest, YbCo$_2$Ge$_4$, is a new stoichiometric Yb-based paramagnetic Kondo lattice~\cite{Kitagawa}. It crystallizes in an orthorhombic structure with two differing Ge sites but only one Co and Yb atomic position, respectively. Clean single crystals, with low residual resistivity $\rho_0=2.4~\mu\Omega$cm display NFL behavior such as a $-\ln T$ dependence of the specific heat coefficient $C/T$ between 0.4 and 10~K, enhancement of the NQR spin-lattice relaxation rate $(T_1T)^{-1}$ upon cooling below 100~K, and a $(\rho-\rho_0) \sim T^{1.4}$ dependence between 0.1 and 1.4~K at ambient pressure and zero field~\cite{Kitagawa}. These NFL effects have been interpreted in terms of a QCP without necessity of fine-tuning. Interestingly, the effective moment estimated from the magnetic susceptibility between 30 and 200~K is $\mu_{\rm eff}=5.5$ $\mu_{\rm B}$, which is larger than the isotropic value for Yb$^{3+}$ of 4.54 $\mu_{\rm B}$, indicating a nearly pure Ising crystalline electric field ground doublet $|J_z \sim \pm 7/2 \rangle$ separated by 200~K from the first excited state and thus an almost trivalent valency of Yb \cite{Kitagawa}. The bulk magnetic susceptibility indicates that the orthorhombic $b$ axis is the magnetic easy direction while $c$ and $a$ are both hard directions. The low-$T$ magnetic anisotropy amounts to $\sim 10$.

A small (0.5 mg mass) high-quality single crystal with improved residual resistivity ratio (RRR) $\rho(300\ {\rm K})/\rho (0\ {\rm K}) \sim 40$, corresponding to $\rho_0=1.8$ $\mu\Omega$cm was grown by the flux method as described in \cite{Kitagawa}. The magnetic Gr\"{u}neisen parameter was directly obtained from the magnetocaloric effect measurement in quasi-adiabatic conditions, $\Gamma_{\rm mag} =T^{-1}(dT/dB)_S$, by utilizing the alternating field technique in a dilution refrigerator~\cite{MCEteq}. For details see the supplemental material (SM)~\cite{SM}. The low-$T$ specific heat was measured utilizing both the relaxation and the quasi-adiabatic heat-pulse technique and complemented by data taken in the PPMS at elevated $T$.

\begin{figure}[t]
\begin{center}
\includegraphics[keepaspectratio, scale=0.6]{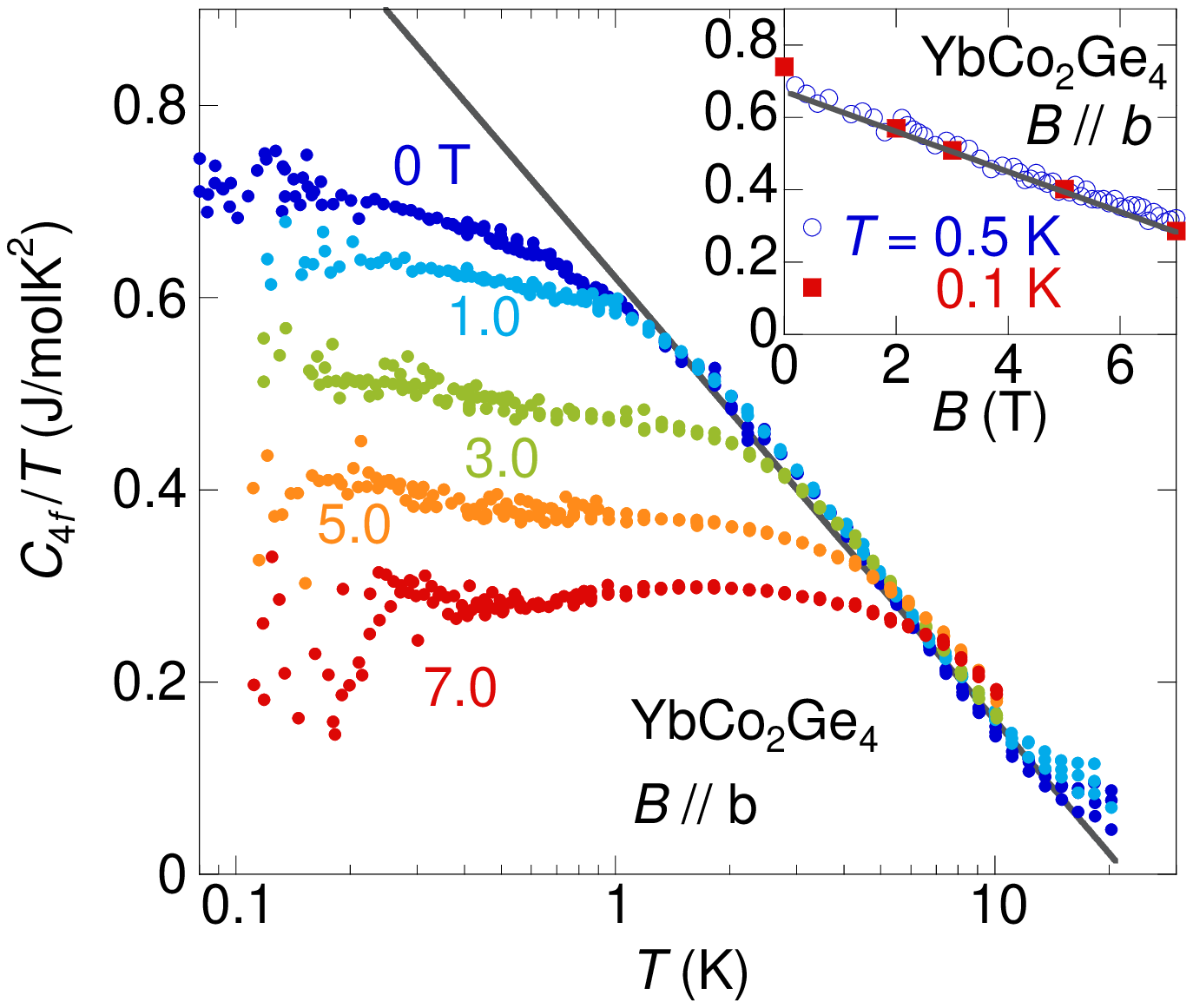}
\caption{(color online) Temperature dependence of the $4f$-electronic contribution of the specific heat divided by temperature, $C_{4f}/T$, at different fields $B\parallel b$. $C_{4f}$ is obtained after subtracting the nuclear contribution $C_{\rm n}\sim 1/T^2$ and phonon contribution estimated from the measurement of the reference material LuCo$_2$Ge$_4$~\cite{SM}. The solid line indicates $C_{4f}/T \sim\ln (T_0/T)$, where $T_0= 22$~K. The inset displays the magnetic field dependence of $C_{4f}/T$ at 0.12 and 0.5~K.}\label{fig1}
\end{center}
\end{figure}

We first focus on the low-temperature heat capacity. Figure \ref{fig1} shows the $4f$-part of the electronic specific heat coefficient after subtraction of nuclear and phonon contributions, cf. SM~\cite{SM}. While a logarithmic divergence at zero field was observed down to $\sim 0.4$ K at $B=0$ previously~\cite{Kitagawa}, our result shows a clear deviation from $-\ln T$ below 1~K. This difference may be related to an improvement of the sample quality. A downward convex shape of $C/T$ versus $\ln T$ has often been ascribed to NFL behavior at three-dimensional antiferromagnetic (AF) QCPs. In this case, $C/T= \gamma -c\sqrt{T}$, as shown by Hertz, Millis, and Moriya~\cite{JAHertz1976,AJMills1993,TMoriya1995}. However, as will be discussed later, this scenario would be incompatible with the temperature dependence of the magnetic Gr\"uneisen parameter and $-dM/dT$. Therefore, we attribute the saturation of the heat capacity coefficient to a crossover from NFL to Fermi liquid (FL) behavior. With increasing magnetic field, this crossover is shifted to higher temperatures. Similar non-diverging behavior of $C_{4f}/T$ is also observed in the field dependence plotted in the inset of Fig. \ref{fig1}. $C_{4f}/T$ increases almost linearly with decreasing field in a wide field region as indicated by the solid line, except for a very small additional enhancement around zero field at $T=0.1$ K. Importantly, the heat capacity data indicate a saturation of the quasiparticle mass, incompatible with zero-field quantum criticality.

\begin{figure}[t]
\begin{center}
\includegraphics[keepaspectratio, scale=0.63]{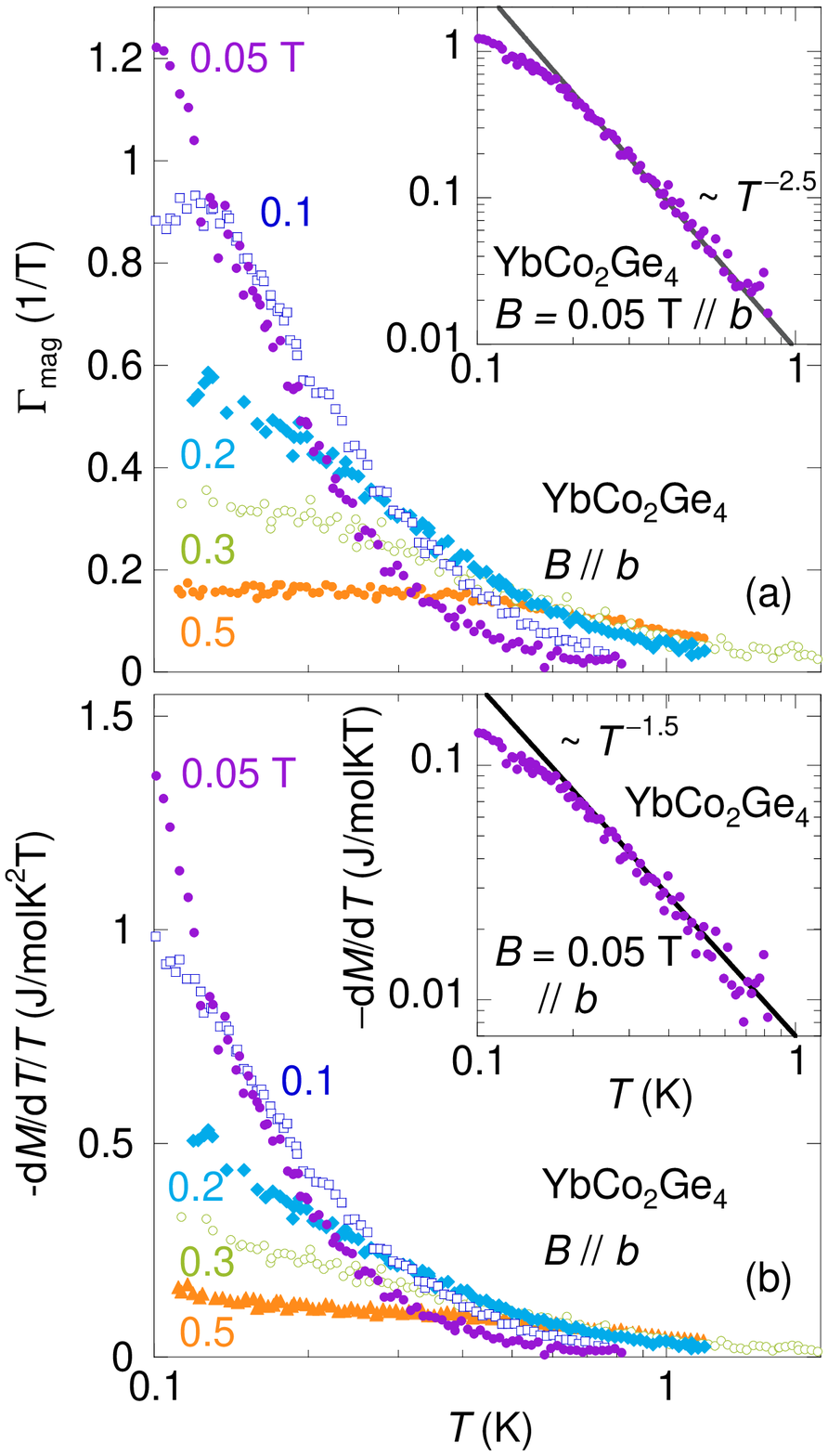}
\caption{(color online) Temperature dependence of the magnetic Gr\"uneisen parameter $\Gamma_{\rm mag}$ (a) and $-(dM/dT)/T=\Gamma_{\rm mag} C/T$ calculated using the heat capacity (b) under various magnetic fields applied parallel to $b$ axis versus a logarithmic temperature scale. The inset in (a) shows a double-log plot of $\Gamma_{\rm mag}$ vs $T$ for $B=0.05$~T. The solid line represents a $T^{-2.5}$ dependence. The inset in (b) displays the respective $-dM/dT$ behavior. The solid line indicates a $T^{-1.5}$ dependence.}\label{fig2}
\end{center}
\end{figure}

We now turn to the magnetic Gr\"uneisen parameter. Figure \ref{fig2} (a) shows the temperature dependence of $\Gamma_{\rm mag}$ at various magnetic fields along the $b$ axis. At $B=0.05$~T, $\Gamma_{\rm mag}(T)$ diverges down to the lowest temperature in contrast with the saturation of $C_{4f}/T$ at $B=0$. With increasing field,  $\Gamma_{\rm mag}$ is suppressed and saturates upon cooling. At the lowest measured field of $B=0.05$~T a strong divergence is found upon upon cooling from 1~K (cf. the inset of Fig. \ref{fig2}). The deviation found below 0.3~K is in accordance with the observed $T/B$ scaling (see below). Interestingly, a similar strong $T^{-2}$ divergence of the magnetic Gr\"uneisen parameter has been found for YbRh$_2$Si$_2$ at $B_{\rm c}=0.06$ T for temperatures above 0.3~K and associated with FM fluctuations~\cite{MCEYbRh2Si2}. In order to study the possible effect of FM fluctuations in YbCo$_2$Ge$_4$ we now concentrate on the temperature derivative of the magnetization, which can be calculated using $-dM/dT= \Gamma_{\rm mag} C$ from the data of the heat capacity and magnetic Gr\"uneisen parameter.

Figure \ref{fig2} (b) shows the temperature dependence of $-(dM/dT)/T$ at $B \parallel b$. Qualitatively similar behavior as for $\Gamma_{\rm mag} (T)$ is found. We focus on the data at 0.05~T displayed in the inset of Fig. \ref{fig2} (b). As indicated by the solid line, $-dM/dT \sim T^{-1.5}$, down to 0.2~K. This is a remarkably strong divergence. It implies a $T^{-0.5}$ divergence of the magnetization at small fields, similar as in $\beta$-YbAlB$_{4}$ \cite{Matsumoto}. The same exponent ($-0.5$) was also found for the differential susceptibility $\chi(T)$ in Au$_{51}$Al$_{34}$Yb$_{15}$~\cite{Deguchi} and the Knight shift in YbRh$_2$Si$_2$~\cite{KIshida2002} and a rather similar one ($-0.7$) for $\chi(T)$ in YbRh$_2$(Si$_{0.95}$Ge$_{0.05}$)$_2$ \cite{PGegenwart2005}. Such divergences in the $\vector{q}=0$ susceptibility indictate the presence of FM fluctuations.

\begin{figure}[t]
\begin{center}
\includegraphics[keepaspectratio, scale=0.7]{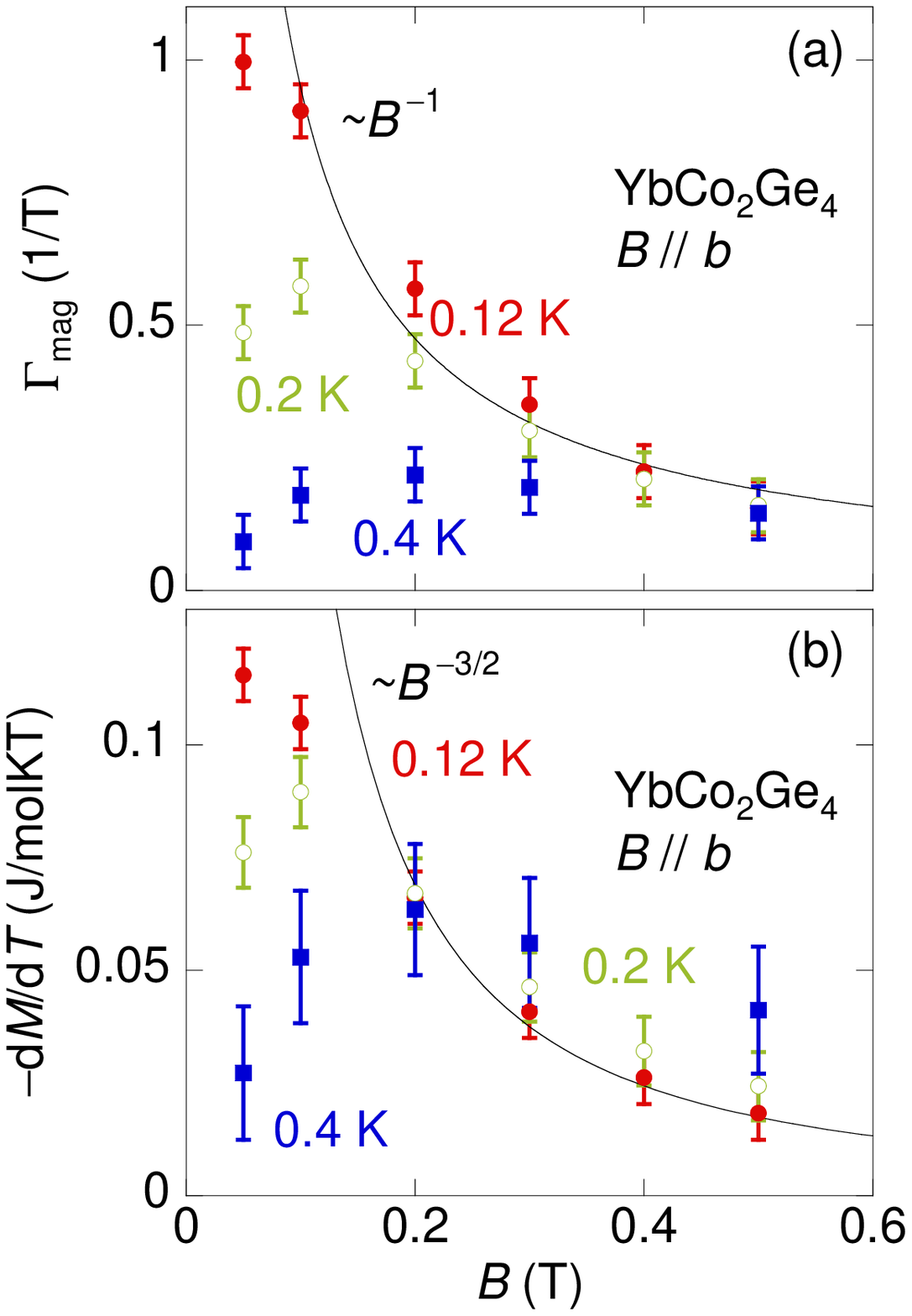}
\caption{(color online) Magnetic field dependence of $\Gamma_{\rm mag}$ (a) and  $-dM/dT$ (b) for YbCo$_2$Ge$_4$ at $B\parallel b$ at various differing temperatures. The solid lines in (a) and (b) indicate $B^{-1}$ and $B^{-3/2}$ dependences, respectively. }\label{fig3}
\end{center}
\end{figure}

Figure \ref{fig3} shows the isothermal field dependence of the magnetic Gr\"uneisen parameter (a) and temperature derivative of the magnetization (b). The solid lines indicate the magnetic field dependences at asymptotically low temperatures as deduces from scaling behavior discussed below. These are $\Gamma_{\rm mag}=(0.095 \pm 0.01)$T$^{-1}B^{-1}$ and $-dM/dT\sim B^{-3/2}$, respectively. Note, that the deviation of the data at 0.12~K at low magnetic fields is caused by the finite temperature effect and fully consistent with $T/B$ scaling behavior discussed next. Similar exponents for $\Gamma_{\rm mag}$ and $-dM/dT$ were found in $\beta$-YbAlB$_{4}$ \cite{Matsumoto} and Au$_{51}$Al$_{34}$Yb$_{15}$~\cite{Deguchi}. For YbRh$_2$(Si$_{0.95}$Ge$_{0.05}$)$_2$ a slightly different behavior, $-dM/dT\sim B^{-4/3}$, has been observed~\cite{PGegenwart2005PhysicaB}.

\begin{figure}[t]
\begin{center}
\includegraphics[keepaspectratio, scale=0.76]{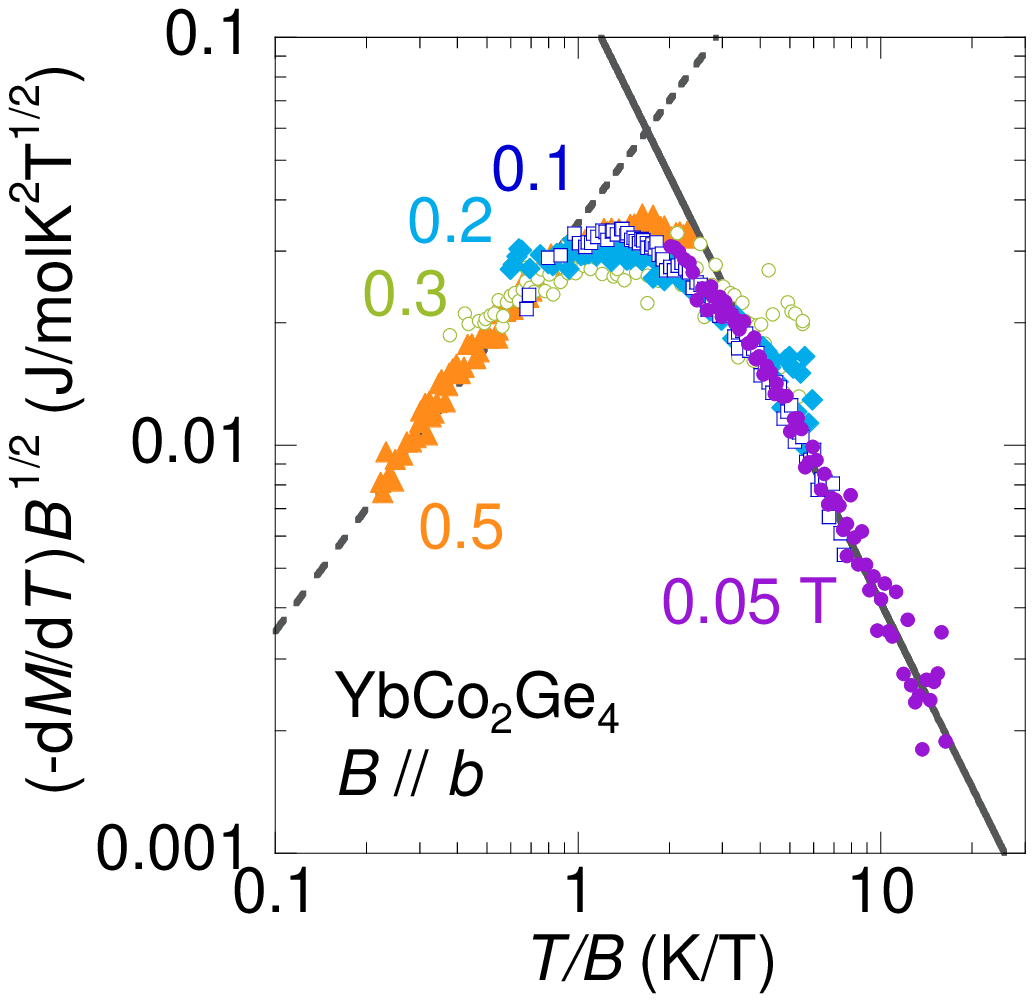}
\caption{(color online) Scaling plot for the temperature derivative of the magnetization $-dM/dT$, measured at various different magnetic fields. The solid and broken lines represent $-dM/dT\propto T^{-3/2}B$ and $-dM/dT\propto TB^{-3/2}$, respectively.}\label{fig4}
\end{center}
\end{figure}

For $\beta$-YbAlB$_{4}$ a characteristic $T/B$ scaling in the temperature derivative of the magnetization has been observed, which was interpreted as signature of the zero-field QCP, since there is no critical field subtracted from the applied field~\cite{Matsumoto}. In Figure \ref{fig4}, we show that our data for YbCo$_2$Ge$_4$ display the very same type of scaling. The lines indicate the asymptotic temperature dependences in the limits of small $T/B$ (FL regime) and large $T/B$ (NFL regime), where $-dM/dT\propto TB^{-3/2}$ (broken line) and $-dM/dT\propto T^{-3/2}B$ (solid line), respectively. The measured data, without exception, collapse on a single curve, spanning two orders
of magnitude in $T/B$. This scaling involves the cross-over between the FL and NFL regimes in wide regions of phase space where the Zeeman energy plays the role of upper boundary of the FL regime~\cite{Matsumoto}.
%The observed linear relation between temperature and field is thus characteristic for the polarization of FM fluctuations.

For $\beta$-YbAlB$_{4}$ such scaling has been interpreted in terms of a FL breakdown due to unconventional quantum criticality~\cite{Matsumoto}. However, the clear lack of a quasiparticle mass divergence in YbCo$_2$Ge$_4$ excludes this possibility for the latter material. Furthermore, it has been proposed, that zero-field quantum criticality in $\beta$-YbAlB$_{4}$ is not accidental, but rather indicates a quantum critical state or (pressure insensitive) quantum critical phase~\cite{Matsumoto}. Hydrostatic pressure experiments of the electrical resistivity have shown that long-range AF ordering sets in only at pressures beyond 2.5~GPa while at low pressures a NFL behavior is realized~\cite{Tomita2015}. Similar observations have also been made in case of Au$_{51}$Al$_{34}$Yb$_{15}$ quasicrystal~\cite{Deguchi}. For $\beta$-YbAlB$_{4}$ an intermediate valence state with Yb$^{+2.75}$ has been found by X-ray photoemission spectroscopy at 25~K \cite{Okawa2010} and a mixed valence state has been deduced from the bulk magnetic properties of Au$_{51}$Al$_{34}$Yb$_{15}$ quasicrystal as well~\cite{Deguchi}. The NFL effects on the latter system have been proposed to arise from quantum valence criticality \cite{Watanabe2010}. However, it seems unlikely, that a stoichiometric material is naturally located close to such a special point in multiparameter phase space. For YbCo$_2$Ge$_4$, quantum valence critical scenarios can be excluded from the large effective moment observed in the Curie-Weiss behavior of the magnetic susceptibility, which proves a stable Yb$^{3+}$ state in this material~\cite{Kitagawa}. 

The $T/B$ scaling and divergence of the magnetic Gr\"uneisen parameter taken together would suggest a zero-field FM QCP in YbCo$_2$Ge$_4$. However, the $C/T$ data at low temperatures are clearly incompatible with the theoretical expectation of a logarithmic or even weak power-law divergence for 3D or 2D FM QCPs, respectively~\cite{JAHertz1976,AJMills1993,TMoriya1995}. Therefore, we need to exclude quantum criticality as origin of the observed $T/B$ scaling. Previous $^{59}$Co NMR/NQR measurements on YbCo$_2$Ge$_4$ have found that the spin-lattice relaxation rate $1/T_1T$ increases upon cooling from high temperatures and bends over around 1~K~\cite{Kitagawa}. As shown in SM~\cite{SM}, the scattering of the data is quite large and does not allow to distinguish a true saturation from a weak power-law divergence suggested by our $dM/dT$ analysis. Since $1/T_1T$ is proportional to the $\vector{q}$ averaged susceptibility, disparate behavior betweeen the spin lattice relaxation rate and the bulk susceptibility may also signal competing finite $\vector{q}$ and $\vector{q}=0$ fluctuations~\cite{KIshida2002}. Another possibility is that a small concentration of paramagnetic impurities, being hardly visible in our heat capacity and the previous NQR measurements (which both were done on crystals of the same batch with very good RRR~$\sim 40$), has a strong influence on the temperature dependence of the magnetization and the magnetic Gr\"uneisen parameter for Kondo lattices with strong FM correlations.

More generally, our results thus indicate, that $T/B$ scaling and divergent magnetic Gr\"uneisen parameter are insufficient for proving a zero-field QCP and confirmation by other properties is required. For geometrically frustrated CeRhSn, a zero-field QCP has indeed been demonstrated by an additional divergence of the (thermal) Gr\"uneisen ratio $\Gamma\sim\alpha/C$ of thermal expansion to specific heat~\cite{Tokiwa2015}. Unfortunately, crystals of YbCo$_2$Ge$_4$, similar as for $\beta$-YbAlB$_{4}$, are much too small for high-resolution thermal expansion measurements. From electrical resistivity, displaying power-law behavior with exponent below 2, it is typically difficult to unambiguously prove that a material is located exactly at a QCP. Previous measurements on YbCo$_2$Ge$_4$ have been described by a NFL $\Delta \rho \sim T^{1.4}$ dependence between 0.1 and 1.4~K~\cite{Kitagawa}, however, it is impossible to exclude a crossover to $T^2$ behavior below 0.2~K. Preliminary hydrostatic pressure experiments indicate that long-range ordering is observed above 3~GPa at $T > 2$~K, supporting our view, that the material is located close to, but not directly at a QCP.

To summarize, the magnetic Gr\"uneisen parameter and temperature derivative of the magnetization in YbCo$_2$Ge$_4$ indicate $T/B$ scaling with zero critical field. Since the heat capacity coefficient saturates, however, zero-field quantum criticality is excluded. Interestingly, similar behavior in the magnetic Gr\"uneisen parameter has previously been found in several other Yb-based heavy-fermion metals. Our study indicates, that $T/B$ scaling and a respective divergence of the magnetic Gr\"uneisen parameter cannot be taken as direct evidence for a zero-field QCP. 

We thank M. Garst and Y. Uwatoko for the useful discussions. This work has been partially supported in Japan by projects KAKENHI No. 15K13523 and 26707018, and in Germany by the JSPS program for Postdoctoral Fellow Research Abroad. P.G. acknowledges the hospitality of the Aspen Center for Physics (NSF grant number 1066293).

\bibliographystyle{apsrev4-1}
\bibliography{bibfile_YbCo2Ge4}

\vspace{1 cm}
\textbf{SUPPLEMENTAL MATERIAL}
\vspace{1 cm}

\begin{figure}[b]
\begin{center}
\includegraphics[keepaspectratio, scale=0.88]{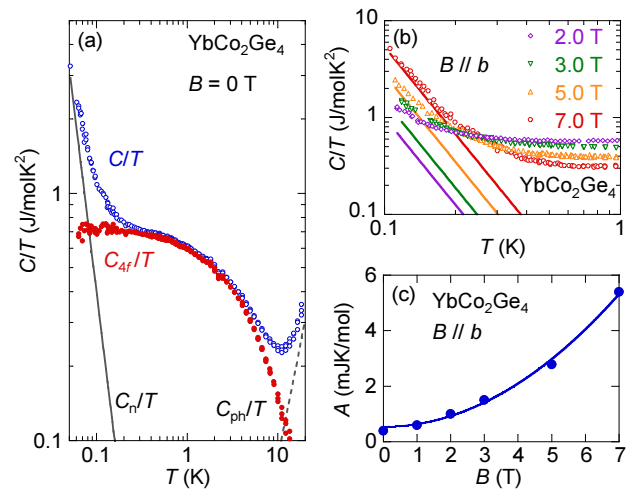}
\caption{(a): Temperature dependence of the specific heat coefficient of 
YbCo$_2$Ge$_4$. Blue circles indicate raw data, red filled circles the 4$f$ contribution after subtraction of the phonon part (dashed line) and the nuclear part (solid line). (b): Respective data at finite magnetic fields $B\parallel b$ with the lines indicating the nuclear contributions. (c) $B$ dependence of the coefficient of the nuclear contribution $C_{\rm n}=A/T^2$. }\label{fig1}
\end{center}
\end{figure}

\begin{table}[b]
\caption{Parameters for the adiabatic magnetocaloric effect measurements. $f$ and $\Delta B$ denote the used frequency and amplitude of the field modulation in the respective temperature range.}
\begin{tabular}[b]{|c|c|c|}
\hline
Temperature range (K) & $f$ (Hz) & $\Delta B$ (mT)\\
\hline
0.1 - 0.3 & 0.03 & 5.5 \\
\hline
0.3 - 0.8 & 0.04 & 6.4 \\
\hline
0.8 - 2.0 & 0.05 & 7.3\\
\hline
\end{tabular}\label{table1}
\end{table}

\begin{figure}[t]
\begin{center}
\includegraphics[keepaspectratio, scale=0.6]{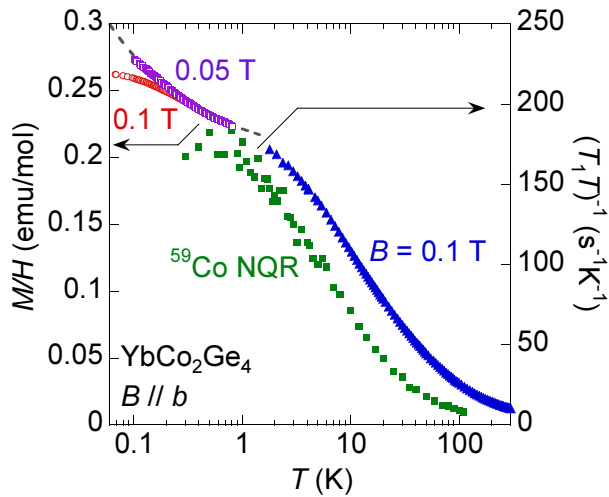}
\caption{Left axis: $T$ dependence of $M/H$ obtained by integrating $dM/dT=-\Gamma_{\rm mag} C$ at $B = 0.05$ T (open square) and 0.1 T (open circle), and  $M/H$ measurement above 2 K at $B = 0.1$ T  (solid triangle)~\cite{Kitagawa}. The broken line indicates $M/H=c_0+c_1/T^{0.5}$ behavior. For all data $B\parallel b$. Right axis: $T$ dependence of the $^{59}$Co-NQR relaxation rate $(T_1T)^{-1}$ (solid square)~\cite{Kitagawa}. }\label{fig2}
\end{center}
\end{figure}

\section{Details on the adiabatic magnetocaloric effect measurements}
%\vspace{1 cm}

We utilized an alternating field technique as described in \cite{MCEteq}. A sinusoidal field modulation with amplitude $\Delta B$ and frequency $f$ around the average field $B$ is used. Both parameters need be chosen such that the magnetic Gr\"uneisen parameter $\Gamma_{\rm mag}=T^{-1}\Delta T/\Delta B$ is independent of $\Delta B$ and $f$, as required in adiabatic conditions. 
The used values are summarized in Tab. 1.

%\vspace{1 cm}
\section{Estimation of $4f$-electron contribution of specific heat $C_{4f}$}
%\vspace{1 cm}

Figure 5(a) shows the zero-field raw data together with the 4$f$ contribution. The phonon contribution has been obtained from a measurement of the reference material LuCo$_2$Ge$_4$. It amounts to $C_{\rm ph}/T=\gamma_0+a_1 T^2$, where $\gamma_0 = 12.9$ (mJ/molK$^2$) and $a_1 =0.713$ (mJ/molK$^4$).  The nuclear contribution $C_{\rm n}$ (solid line) is estimated from the data in finite magnetic fields, for which $C_{\rm n}(T,B)=A(B)/T^2$ is first determined at high field as shown in Fig. 5 (b) and then extrapolated to low fields ($B =$ 0 T and 1 T), assuming a $B^2$ dependence of $A(B)$, as shown by the solid line in Fig. 5 (c). The validity of the value at zero field $A(B= 0 {\rm T})\sim 4\times 10^{-4}$ (JK/mol) can be also confirmed by the nuclear quadrupole $Q$ and electric field $q$, namely $C_{\rm n}T^2=\sum_{i} \frac{R}{45}(\nu_Q h/k_{\rm B})^2I(I+1)(2I+1)(2I-1)r_i/100$, where $R =$ gas constant, $h=$ Planck constant, $k_{\rm B}=$ Boltzmann constant, $I=$ nuclear spin, $\nu_Q=3eqQ/4I(2I-1)$ NQR frequency and $r_i=$ natural abundance \cite{Bleaney1963}. The contribution from the $^{59}$Co ($I=7/2$, $r=100$ \% ) is calculated from the NQR result $\nu_Q=1.5$ MHz \cite{Kitagawa}. $C_{\rm n}T^2=1.8 \times 10^{-6}$ (JK/mol) is two orders of magnitude smaller than the experimental value. In Yb-based compounds, nuclear contribution at $B=0$ is usually dominated by the $^{173}$Yb ($I=5/2$, $r=16.13$ \%) nuclei due to the large quadrupole moment. If we assume $A(B= 0 {\rm T})\sim 4\times 10^{-4}$ (JK/mol) arises only from $^{173}$Yb nuclei, $\nu_Q\sim150$ MHz is necessary. In fact, this is consistent with the value calculated from the point charge model for $^{173}$Yb nuclei $\nu_Q\sim130$ MHz, assuming the antishielding factor $\gamma_{\infty}=-50$ \cite{Edvardsson1998}.

%\vspace{1 cm}
\section{Comparison of magnetic susceptibility and the $^{59}$Co-NQR spin lattice relaxation rate}
%\vspace{1 cm}

Figure 6 compares the $1/T_1T$ data derived from $^{59}$Co-NQR~\cite{Kitagawa} with the magnetic susceptibility. For the latter, we used published data down to 2~K. In addition for very low temperatures data were obtained by integration of $dM/dT=-\Gamma_{\rm mag} C$ where the integraton constant has been adjusted to achive a smooth evolution of $M(T)$. The NQR relaxation rate divided by $T$ follows upon cooling from 100 K qualitatively the temperature dependence of the magnetization, though there is no simple proportionality. Below 1 K, the noise level of $1/T_1T$ is unfortunately too high to distinguish a saturation from a weak power law increase. The increase of $M(T)$ between 0.4 and 1K amounts to $\sim6 \%$ only, while the noise level of $1/T_1T$ in the same temperature range is about $13 \%$. Therefore, it is not clear if the temperature dependence of the relaxation rate is disparate from that of the magnetization. Implications are discussed in the main part of the manuscript.

%\bibliographystyle{apsrev4-1}
%\bibliography{bibfile_YbCo2Ge4_suppl}

\end{document}